\documentclass[conference]{IEEEtran}
\IEEEoverridecommandlockouts

\usepackage{amsmath,amssymb,amsfonts}
\usepackage{graphicx}
\usepackage{textcomp}
\usepackage{xcolor}
\usepackage{epsfig,epstopdf}

\usepackage{amsopn, fancybox, color, bm, dsfont , latexsym, accents, subfigure, algorithm, algpseudocode, multirow, caption}
\usepackage{pifont, wasysym, stmaryrd}
\usepackage[hidelinks]{hyperref}

\def\BibTeX{{\rm B\kern-.05em{\sc i\kern-.025em b}\kern-.08em
    T\kern-.1667em\lower.7ex\hbox{E}\kern-.125emX}}

\begin{document}

\title{Enhanced Informed Dynamic BP Decoding
Scheduling Strategies for 5G NR LDPC Codes%
\thanks{This is the author's accepted manuscript
of a paper published in the \emph{2022 IEEE 96th
Vehicular Technology Conference (VTC2022-Fall)}.
The final published version is available at
\url{https://doi.org/10.1109/VTC2022-Fall57202.2022.10013051}.}
\thanks{© 2022 IEEE. Personal use of this material is permitted.
Permission from IEEE must be obtained for all other uses,
in any current or future media,
including reprinting/republishing this material for advertising or promotional purposes,
creating new collective works,
for resale or redistribution to servers or lists,
or reuse of any copyrighted component of this work in other works.}
}

\makeatletter
\newcommand{\linebreakand}{%
  \end{@IEEEauthorhalign}
  \hfill\mbox{}\par
  \mbox{}\hfill\begin{@IEEEauthorhalign}
}
\makeatother

\author
{
  \IEEEauthorblockN{Tofar C.-Y. Chang}
  \IEEEauthorblockA{\textit{Dept. of Electrical Eng.} \\
    \textit{National Taipei University of Technology}\\
    Taipei, TAIWAN \\
    tofar@ntut.edu.tw}
  \and
  \IEEEauthorblockN{I-Hsiang Lee}
  \IEEEauthorblockA{\textit{MediaTek Inc.} \\
    Hsinchu, TAIWAN \\
    i-hsiang.lee@mediatek.com}
  \and
  \IEEEauthorblockN{Pin-Han Wang}
  \IEEEauthorblockA{\textit{MediaTek Inc.} \\
    Hsinchu, TAIWAN \\
    Pin-han.Wang@mediatek.com}
  \linebreakand 

  \IEEEauthorblockN{Jian-Jia Weng}
  \IEEEauthorblockA{\textit{Dept. of Electrical Eng.} \\
    \textit{National Taiwan Ocean University}\\
    Keelung, TAIWAN \\
    jjweng@mail.ntou.edu.tw}

  \and
  \IEEEauthorblockN{Yu T. Su}
  \IEEEauthorblockA{\textit{Dept. of Electrical and Computer Eng.} \\
    \textit{National Yang Ming Chiao Tung University}\\
    Hsinchu, TAIWAN \\
    ytsu@nycu.edu.tw}
}

\maketitle
\begin{abstract}
We present an enhanced informed dynamic scheduling (IDS) strategy to avoid propagations of unreliable messages in the residual belief-propagation (RBP) algorithm for decoding low-density parity-check (LDPC) codes. To reduce the resource needed for updating the degree-one variable nodes (VNs) in an RBP-based algorithm for decoding 5G New Radio (NR) LDPC codes, we further introduce a process consisting of several simple procedures for managing the resource consumption. Simulation results demonstrate that our RBP algorithm based on the proposed IDS strategy outperforms existing RBP-based algorithms at the early decoding iterations. By using our degree-one VN resource managing approach, the performance of both the proposed and the existing decoders is significantly improved in decoding the 5G NR codes.
\end{abstract}


\section{Introduction}
It has been shown that
low-density parity-check (LDPC) codes provide capacity-approaching performance when the belief propagation (BP) algorithm is used for decoding.
Due to their excellent error-correcting capability, LDPC codes have been considered as the channel codes in several wireless communication standards such as IEEE 802.11ac (WiFi) \cite{WiFi}, IEEE 802.16e (WiMAX) \cite{WiMax}, and the 3GPP 5G New Radio (NR) specification \cite{NR SPEC}.

For the conventional BP algorithm, all the variable-to-check (V2C) messages are updated simultaneously, so are all check-to-variable (C2V) messages propagated. Such a fully-parallel updating scheme, which is called \emph{flooding} scheduling, often need many decoding iterations to achieve the required error-rate performance. To have better decoding efficiency and performance, the standard sequential scheduling (SSS) \cite{LBP}
and informed dynamic scheduling (IDS) strategies have been introduced. It has been shown that some IDS-assisted BP algorithms do yield impressive performance improvement against those using the SSS or flooding scheduling.


The first IDS-aided BP decoder, the residual belief propagation (RBP) algorithm, determines the updating order of the C2V messages based on
the C2V \emph{residuals}, which is defined as the difference between the previous C2V message value and its precomputed update value. It
offers impressive performance at the first few iterations but fails to yield satisfactory converged performance. This might due to the fact that
the RBP decoding process tends to focus on a small part of nodes/edges of the code graph and thus deprive some erroneous decisions of the
opportunities to be corrected. To remedy this greedy behavior in RBP decoding, several related modified decoding strategies have been proposed in the literature \cite{RBP}--\cite{IDS21}.
Among these improved (or we called RBP-based) algorithms, 
the silent-variable-node-free (SVNF) RBP algorithm introduced in \cite{SVNF}
can not only yield quite nice performance but provide almost best performance-complexity tradeoff.

Most IDS strategies, as just mentioned, aim at reducing the negative effects brought about by the greedy approach and encouraging the propagation of reliable information. On the other hand, we must also pay attention to the cases when erroneous messages were spread. Our extensive computer simulations reveal that sometimes the unreliable messages can be spread via the dynamic scheduling used. It is therefore necessary that an IDS strategy can also prevent or suspend the propagation of unreliable messages.

In each update, the RBP algorithm forwards the C2V message with the maximum residual. A check node (CN) may send messages to its connected variable nodes (VNs) several times but between two consecutive C2V messages sent by the same CN, there might be no new information received by this CN. As no updated V2C message is received, the two C2V messages are generated from the same set of V2C messages. A large C2V residual sometimes implies that the associated VN or the related V2C message might carry unreliable information \cite{IDS16a}. Hence, if the C2V message with the maximum residual is sent by the associated CN, this CN should stop delivering new C2V messages until it obtains new V2C information to avoid forwarding V2C information.

In this paper, we propose what we call the \emph{update-before-propagate RBP (UBPRBP)} algorithm. After a CN sends its C2V message in UBPRBP algorithm, this CN will temporarily stay \emph{muted} until it can receive new information from its connecting VNs; no C2V message can be delivered as the associated CN is set muted. Simulation results show that compared with the RBP algorithm, our algorithm can provide better decoding performance
with lower computational complexity.

Furthermore, we found that the LDPC codes used in the 5G NR specification contain many degree-1 VNs. It is observed that the C2V messages associated with the degree-1 VNs are more likely to be updated than the other edges if an IDS decoder is applied.
However, this would reduce the decoding efficiency since a degree-1 VN can only provide its intrinsic channel information to the connected CN.
As a result, we propose a scheme to remedy this shortcoming. With our method, all degree-1 VN can only be updated at the end of the iteration;
when an edge linking to a degree-1 VN is selected, the passing of the C2V message will be performed by one of the other edges linking to the same CN instead. Simulation results show that our degree-1 VN handling process would significantly improve the performance of the RBP-based algorithms
when decoding the NR LDPC codes.

The rest of this paper is organized as follows. A brief review of the RBP algorithm is given in Section II, and the UBPRBP algorithm is also presented in the same section. Section III illustrates the proposed degree-1 VN handling process. Simulation results are shown in Section IV,
and the conclusion remarks are drawn in Section V. 
\section{The Proposed UBPRBP Algorithm}
A binary ($N$, $K$) LDPC code $\mathcal{C}$ of rate $R = K/N$ is defined by
an $M \times N$ parity-check matrix $\bm{H}=[H_{mn}]$,
where the columns and rows respectively represent the VNs and CNs
on the associated Tanner graph and $M$ is the number of rows in the $\bm{H}$.
It is assumed that
the BPSK modulation is used and
a binary codeword $\bm{u}=(u_0,u_1,\cdots,u_{N-1})$ is mapped to
the transmitted sequence $\bm{x}=(x_0,x_1,\cdots,x_{N-1})$
where $x_n = 1-2 u_n$ for $0\leq n < N$
and transmitted over a zero-mean AWGN channel with noise variance $\sigma^2$.
We also denote the corresponding received noisy codeword
and the decoded decision vector as
$\bm{y}=(y_0,y_1,\cdots,y_{N-1})$ and $\hat{\bm{u}}=(\hat{u}_0,\hat{u}_1,\cdots,\hat{u}_{N-1})$, respectively.

Let $L_{m\rightarrow n}^{\text{c2v}}$ be the C2V message from $c_m$, the $m$th CN, to $v_n$, the $n$th VN,
$L_{n\rightarrow m}^{\text{v2c}}$ present the V2C message from $v_n$ to $c_m$,
and $L_n$ be the total log-likelihood ratio (LLR) of $v_n$.
We define $\mathcal{M}(n)=\{m|H_{mn}=1\}$ as the index set of CNs linked to $v_n$
and $\mathcal{N}(m)=\{n|H_{mn}=1\}$ as the index set of VNs connected to $c_m$.
We further denote the set $\mathcal{M}(n)$ with $c_m$ excluded by $\mathcal{M}(n)\setminus m$
and similarly the set $\mathcal{N}(m)$ with $v_n$ excluded by $\mathcal{N}(m)\setminus n$.
For BP decoding algorithms,
the V2C messages sent from $v_n$ to $c_m$, where $m\in \mathcal{M}(n)$, can be computed by
\begin{eqnarray}\label{eqn:V2C}
  L_{n\rightarrow m}^{\text{v2c}} &=& \frac{2y_n}{\sigma^2} + \sum_{m'\in\mathcal{M}(n)\setminus m} L_{m'\rightarrow n}^{\text{c2v}},
\end{eqnarray}
and the C2V messages delivered from $c_m$ to $v_n$ for each $n\in \mathcal{N}(m)$ are updated by
\begin{eqnarray}\label{eqn:C2V}
  L_{m\rightarrow n}^{\text{c2v}} &=& 2\tanh^{-1}\left(\prod_{n'\in\mathcal{N}(m)\setminus n} \tanh\left(\frac{1}{2} L_{n'\rightarrow m}^{\text{v2c}}\right) \right).
\end{eqnarray}
The total LLR of $v_n$ is
\begin{eqnarray}\label{eqn:VLLR}
    L_n = \frac{2y_n}{\sigma^2}+\sum_{m\in\mathcal{M}(n)} L_{m\rightarrow n}^{\text{c2v}},
\end{eqnarray}
and the decision is obtained via $\hat{u}_{n}=\text{sgn}(L_n)$,
where $\text{sgn}(\cdot)$ denotes the sign function defined by $\text{sgn}(a)=1$ if $a\geq0$ and $\text{sgn}(a)=-1$ otherwise.
\subsection{The RBP Algorithm}

The RBP algorithm can be viewed as the first IDS-aided decoder
which decides the message updating order by
predicting the possible benefit brought by sending a message.
In the RBP algorithm,
the \emph{residual} of a C2V message,
denoted by $R_{m\rightarrow n}^{\text{c2v}}$,
is defined as the difference between its current and precomputed values:
\begin{eqnarray}\label{eqn:C2VRSD}
  R_{m\rightarrow n}^{\text{c2v}} = |\hat{L}_{m\rightarrow n}^{\text{c2v}}-L_{m\rightarrow n}^{\text{c2v}}|,
\end{eqnarray}
where $\hat{L}_{m\rightarrow n}^{\text{c2v}}$ is the precomputed version of $L_{m\rightarrow n}^{\text{c2v}}$.
The C2V residual is thought of as a metric for measuring how far a message is from convergence \cite{RBP},
and the C2V message with the maximum residual
will be selected for propagation in each message update in the RBP algorithm.

At the beginning of the RBP decoding,
each C2V message ($L_{m\rightarrow n}^{\text{c2v}}$) is initialized to $0$,
and the precomputed C2V message ($\hat{L}_{m\rightarrow n}^{\text{c2v}}$) is generated via (\ref{eqn:C2V}).
Then, the residual of a C2V message $R_{m\rightarrow n}^{\text{c2v}}$
is computed by (\ref{eqn:C2VRSD}).
Denote $c_{m^*}$ and $v_{n^*}$ as the CN and VN corresponding to the message
conveying the maximum residual, respectively.
As $c_{m^*}$ and $v_{n^*}$ are identified,
the C2V message $L_{m^*\rightarrow n^*}^{\text{c2v}}$
would be replaced by $\hat{L}_{m^*\rightarrow n^*}^{\text{c2v}}$ and propagated,
and then the V2C messages $L_{n^*\rightarrow i}^{\text{v2c}}$, $i \in \mathcal{M}(n^*) \setminus m^*$ would thus be updated by (\ref{eqn:V2C}).
Then, $\hat{L}_{i\rightarrow j}^{\text{c2v}}$ and $R_{i\rightarrow j}^{\text{c2v}}$ for all $j \in \mathcal{N}(i)\setminus n^*,~i \in \mathcal{M}(n^*) \setminus m^*$
will be calculated before the next C2V message selection.
The above procedures will be repeated
until the stopping criterion is satisfied, that is,
a valid codeword is obtained/the maximum iteration (or updated edge number) is reached.
Generally, an iteration would be counted after $E$ C2V messages are propagated,
where $E$ is the number of edges on the code graph corresponding to $\bm{H}$.

\subsection{Update-before-Propagate RBP Algorithm}
Recently, the related works have suggested several effective strategies
to prevent the greedy phenomenon
and have reliability messages can be passed as soon as possible during decoding.
However, the prevention of the unreliable message propagation
is rarely addressed in the literature.
Through our experiments, we observed that
a CN can propagate its C2V messages multiple times
before it receives new information from the linked VNs.
As a VN associated with the largest C2V residual
indicates that the decision of this VN is possibly unreliable,
propagating multiple C2V messages from a CN with no V2C information update between the message propagations
would have the unreliable VNs send unreliable messages to the other VNs linked to the same CN.
A simple example is given as follows.
Consider that
an all-zero codeword is sent and we are at the beginning of the decoding,
where all V2C messages are initialized as $2 y_n/\sigma^2$.
Assume a CN, say $c_0$, has three linked VNs, say $v_0$, $v_1$, and $v_2$ respectively,
and only $v_0$ has negative channel reliability value, i.e., $2y_0/\sigma^2<0$.
Suppose the residual $R_{0\rightarrow 0}^{\text{c2v}}$ is the maximum and
thus the corresponding C2V message is selected to propagate for the first message update.
If in one of the future updates,
$R_{0\rightarrow 1}^{\text{c2v}}$ becomes the maximum residual
but $v_0$ has not received C2V messages from its connected CNs except for that from $c_0$,
the incorrect information of $v_0$ would be incorporated in the C2V message $L_{0\rightarrow 1}^{\text{c2v}}$
and pass to $v_1$.

To suppress the potential error propagations illustrated above,
we propose the update-before-propagate RBP (UBPRBP) algorithm.
Let $\mathcal{O}_M=\{0,1,\ldots,M-1\}$ and $\mathcal{O}_N=\{0,1,\ldots,N-1\}$.
Denote $\mathcal{S}$ as the group of unmuted CNs,
and $Q(m)$
is the VN index which corresponds to the largest C2V residual associated with $c_m$.
The UBPRBP algorithm is described in \textbf{Algorithm \ref{UBPRBP_algo}}.

\begin{algorithm}[h]
    \caption{UBPRBP Algorithm}
    \label{UBPRBP_algo}
    \begin{algorithmic}[1]
    \State Initialize all $L_{m\rightarrow n}^{\text{c2v}}=0$ and $L_{n\rightarrow m}^{\text{v2c}}=2y_n/\sigma^2$. Let $\mathcal{S} \leftarrow \mathcal{O}_M $.
    \State Generate all $\hat{L}_{m\rightarrow n}^{\text{c2v}}$ by (\ref{eqn:VLLR}) and compute all $R_{m\rightarrow n}^{\text{c2v}}$ by (\ref{eqn:C2VRSD})

    \State Let $Q(m)\leftarrow\arg\max_{n}\left\{ R_{m\rightarrow n}^{\text{c2v}}|n\in\mathcal{N}(m)\right\}$ $\forall m \in \mathcal{S}$
    \State Find $(m^*,n^*)=\arg\max_{(m,Q(m))}\{R_{m\rightarrow Q(m)}^{\text{c2v}} | m\in \mathcal{S}\}$ and let $L_{m^*\rightarrow n^*}^{\text{c2v}}\leftarrow \hat{L}_{m^*\rightarrow n^*}^{\text{c2v}}$
        \label{ALG:UBPRBP_findmaxRSD}
    \State Propagate $L_{m^*\rightarrow n^*}^{\text{c2v}}$ and set $R_{m^*\rightarrow n^*}^{\text{c2v}}=0$
    \State Let $\mathcal{S} \leftarrow \mathcal{S}\backslash m^*$
    \For{every $i\in\mathcal{M}(n^*)\backslash m^*$}
        \State Generate and propagate $L_{n^*\rightarrow i}^{\text{v2c}}$ and let $\mathcal{S} \leftarrow \mathcal{S} \cup \{i\}$
            \State Compute $\hat{L}_{i\rightarrow j}^{\text{c2v}}$ and update $R_{i\rightarrow j}^{\text{c2v}}~\forall j\in\mathcal{N}(i)\backslash n^*$
        \State Find $Q(i)=\arg\max_{j}\left\{R_{i\rightarrow j}^{\text{c2v}}|j\in\mathcal{N}(i)\right\}$
    \EndFor
    \If{Stopping condition is not satisfied}
        \State Go to line \ref{ALG:UBPRBP_findmaxRSD}
    \EndIf
    \end{algorithmic}
\end{algorithm}

In \textbf{Algorithm \ref{UBPRBP_algo}},
all the largest residuals associated with the unmuted CNs
($R_{m\rightarrow Q(m)}^{\text{c2v}}$ for all $m\in \mathcal{S}$)
would be compared for selecting the maximum one,
and the C2V message corresponding to the chosen maximum residual would be delivered.
A C2V message sent,
the corresponding CN will be temporary muted (removed from $\mathcal{S}$)
until any one of its linked VNs
receives updated information from other connected CNs
and regenerates the V2C message to this CN.
The stopping condition is set as
the maximum iteration is reached or
a valid codeword is obtained at the end of each iteration.

Compared with the conventional RBP algorithm,
the UBPRBP algorithm requires less computation efforts for the residual comparisons.
This is because each CN's associated largest C2V residual
is found in advance (lines 3 and 10 of \textbf{Algorithm \ref{UBPRBP_algo}}) and
only the largest residual values of unmuted CNs (those in $\mathcal{S}$) will be compared for the edge selection (line 4).

\section{Proposed Process for Updating Degree-One VNs for IDS Strategies}

The LDPC codes specified for the 5G NR
have some characteristics which are rarely seen over the LDPC codes used
in the other communication standards.
One of the characteristics is the appearance of degree-1 VNs on the code graph,
which is designed for supporting a higher level of flexibility
for the incremental-redundancy hybrid automatic-repeat-request (IR-HARQ) schemes \cite{NR SPEC}.

Through our simulations,
we found that the degree-1 VNs can be selected and updated
more frequently than the other VNs
when IDS strategies are applied in decoding.
One of the proper explanations for this phenomenon is given as follows.
Based on the density evolution for LDPC codes with BP decoding \cite{DEvol},
we know that the VN LLR magnitudes will expand
as the iteration count increases and so do the V2C messages.
For a degree-1 VN, however,
the V2C message it sends would always be the channel reliability value ($2y_n/\sigma^2$).
Suppose the degree of VN $v_n$ is one and $c_m$ is its only-connected CN.
After the message-passing decoding is performed a few iterations,
the magnitudes of the V2C messages $L^{v2c}_{n'\rightarrow m}$ for $n'\in\mathcal{N}(m)\setminus n$
might rapidly grow while $L^{v2c}_{n\rightarrow m}$ remains unchanged.
Since the magnitude of a C2V message $L^{c2v}_{m\rightarrow a}$ computed by (\ref{eqn:C2V})
is usually dominated by the smallest magnitude of the V2C messages $L^{v2c}_{a'\rightarrow m}, a'\in\mathcal{N}(m)\setminus a$.
It could be predicted
that the C2V messages' magnitude (either for the propagated or precomputed versions)
$L^{c2v}_{m\rightarrow n'},~n'\in\mathcal{N}(m)\setminus n$
would be dominated by $|L^{v2c}_{n\rightarrow m}|=|2y_n/\sigma^2|$ and
hence delivery of these C2V messages would be eliminated
since the associated residuals can be quite small.
By contrast,
the residual of $L^{c2v}_{m\rightarrow n}$ might be larger than C2V messages
due to the continuous updates of $L^{v2c}_{n'\rightarrow m}~n'\in\mathcal{N}(m)\setminus n$.
As a result,
the C2V message related to a degree-1 VN
would have a higher opportunity of being selected and updated
than the other C2V messages linked to the same CN.
Consuming much decoding resource on updating the degree-1 VNs, however,
would result in low decoding efficiency because
these nodes  can only provide its intrinsic channel information to the
connected CN.

For the IDS strategies
we introduce a process to reduce the efficiency loss
caused by the appearance of the degree-1 VNs.
The proposed \emph{degree-1 VN (D1VN) processing scheme} works based on following guidelines:
1) all degree-1 VNs would be updated only once in each iteration,
and the updating is performed at end of each iteration;
2) in one iteration, the C2V messages associated with the degree-1 VNs can participate the message selection
once only;
3) while the C2V message going to propagate is associated with degree-1 VN,
the propagation would be skipped and executed by another C2V message associated with the same CN.
The purpose of the first two guidelines
is to prevent the decoding resource being frequently occupied by certain nodes.
The last one is aimed at maintaining the propagation of new messages
since a large C2V message residual may indicate that
the VNs linked to this CN have received renewed information from other connected CNs
and thus might have crucial update on their reliability.

In \textbf{Algorithm \ref{RBPd1_algo}},
we present the RBP decoding algorithm which is assisted with
our D1VN processing scheme.
As a degree-1 VN is allowed to join the selection for just one time in an iteration,
we use $\mathcal{F}(n)$ to label if a degree-1 VN, $v_n$, has participated in the selection.
$\mathcal{F}(n) =\mathbb{T} $ (True) indicates the C2V message/residual has been considered in the current iteration;
otherwise $\mathcal{F}(n) =  \mathbb{F}$ (False).
$d(n)$ in \textbf{Algorithm \ref{RBPd1_algo}} denotes
the VN degree of $v_n$.

\begin{algorithm}[h]
    \caption{RBP Algorithm with D1VN Processing Scheme}
    \label{RBPd1_algo}
    \begin{algorithmic}[1]
    \State Initialize all $L_{m\rightarrow n}^{\text{c2v}}=0$ and $L_{n\rightarrow m}^{\text{v2c}}=2y_n/\sigma^2$. Set $\mathcal{F}(j)=\mathbb{F} ~ \forall j \in \mathcal{O}_{N}$
    and the edge counter $\eta =0$.
    \State Generate all $\hat{L}_{m\rightarrow n}^{\text{c2v}}$ by (\ref{eqn:VLLR}) and compute all $R_{m\rightarrow n}^{\text{c2v}}$ by (\ref{eqn:C2VRSD})
    \State Find $(m^*,n^*)=\arg\max_{(m,n)} \{ R_{m\rightarrow n}^{\text{c2v}} |  n\in\mathcal{N}(m)\ \text{and}\ \mathcal{F}(n)=\mathbb{F}, m\in\mathcal{O}_M \}$
    \label{ALG:RBP_findmaxRSD}
    \If{$d(n^*)=1$}
        \State Find $n'=\arg\max_{n} \left\{R_{m^*\rightarrow n}^{\text{c2v}}|n\in\mathcal{N}(m^*)\backslash n^* \right\}$
        \State Set $\mathcal{F}(n^*)=\mathbb{T}$ and then $n^*\leftarrow n'$
    \EndIf
    \State  Let $L_{m^*\rightarrow n^*}^{\text{c2v}}\leftarrow \hat{L}_{m^*\rightarrow n^*}^{\text{c2v}}$; propagate $L_{m^*\rightarrow n^*}^{\text{c2v}}$ and set $R_{m^*\rightarrow n^*}^{\text{c2v}}=0$.
    \For{every $i\in\mathcal{M}(n^*)\backslash m^*$}
        \State Generate and propagate $L_{n^*\rightarrow i}^{\text{v2c}}$
        \For{every $j\in\mathcal{N}(i)\backslash n^*$ and $\mathcal{F}(j)=\mathbb{F}$}
            \State Compute $\hat{L}_{i\rightarrow j}^{\text{c2v}}$ and update $R_{i\rightarrow j}^{\text{c2v}}$
        \EndFor
    \EndFor
    \State $\eta \leftarrow \eta + 1$.
    \If{$\eta=E - E_1$}
        \State Set $\mathcal{F}(n)=\mathbb{F}$ for all $n\in \mathcal{O}_N$
        \State Generate and propagate $L_{m\rightarrow n}^\text{c2v}$ and then set $R_{m\rightarrow n}^{\text{c2v}}= 0$, where $m \in \mathcal{M}(n)$, $n\in \mathcal{O}_N$ and $d(n)=1$.
    \EndIf
    \If{Stopping condition is not satisfied}
        \State Go to line \ref{ALG:RBP_findmaxRSD}
    \EndIf
    \end{algorithmic}
\end{algorithm}

For each iteration of \textbf{Algorithm \ref{RBPd1_algo}},
we initialize $\mathcal{F}(n)=\mathbb{F}$ for all VNs,
and $\mathcal{F}(n)$ would always keep unchanged if $d(n)>1$.
For each message selection,
only the C2V messages connecting to ``$\mathbb{F}$"-labeled VNs would be compared,
and the one with the maximum residual will be sent.
However, once a message associated with a degree-1 VN is chosen,
the propagation of this C2V message will be postponed to the end of the current iteration,
and the C2V message which is from the same CN and has the second maximum residual will be delivered instead.
Furthermore,
the corresponding $\mathcal{F}(n)$ would be set as $\mathbb{T}$ and
temporary excluded from the selection until the next iteration begins.
As only $E$ C2V messages can be sent in one iteration,
the above procedures will be repeated $E-E_1$ times,
where $E_1$ is the number of degree-1 VNs.
While propagating $E-E_1$ C2V messages,
it would simultaneously
send the degree-1 VNs the messages provided by their linked CNs
in \textbf{Algorithm \ref{RBPd1_algo}}.


The D1VN processing scheme can also
be adopted to the UBPRBP algorithm (\textbf{Algorithm \ref{UBPRBP_algo}})
for decoding the 5G NR LDPC codes.
The D1VN processing scheme assisted UBPRBP algorithm
is described in \textbf{Algorithm \ref{CNUSd1_algo}}.

\begin{algorithm}[h]
    \caption{UBPRBP with D1VN Processing Scheme}
    \label{CNUSd1_algo}
    \begin{algorithmic}[1]
    \State Initialize all $L_{m\rightarrow n}^{\text{c2v}}=0$ and $L_{n\rightarrow m}^{\text{v2c}}=2y_n/\sigma^2$. Set $\mathcal{F}(j)=\mathbb{F}~\forall j \in \mathcal{O}_{N}$,  $\mathcal{S} = \mathcal{O}_M $ and the edge counter $\eta =0$.
    \State Generate all $\hat{L}_{m\rightarrow n}^{\text{c2v}}$ by (\ref{eqn:VLLR}) and compute all $R_{m\rightarrow n}^{\text{c2v}}$ by (\ref{eqn:C2VRSD})

    \State Let $Q(m)\leftarrow\arg\max_{n}\left\{R_{m\rightarrow n}^{\text{c2v}}|n\in\mathcal{N}(m)\right\}$ $\forall m \in \mathcal{S}$
        \State Find $(m^*,n^*)=\arg\max_{(m,Q(m))}\{R_{m\rightarrow Q(m)}^{\text{c2v}}|\mathcal{F}(Q(m))=\mathbb{F},m\in \mathcal{S}\}$  \label{ALG:UBPRBPd1_findmaxRSD}
    \If{$d(n^*)=1$}
        \State Find $n'=\arg\max_{n}\left\{R_{m^*\rightarrow n}^{\text{c2v}}|n\in\mathcal{N}(m^*)\backslash n^*\right\}$
        \State Set $\mathcal{F}(n^*)=\mathbb{T}$ and then $n^*\leftarrow n'$
    \EndIf
    \State Let $L_{m^*\rightarrow n^*}^{\text{c2v}}\leftarrow \hat{L}_{m^*\rightarrow n^*}^{\text{c2v}}$; propagate $L_{m^*\rightarrow n^*}^{\text{c2v}}$ and let $R_{m^*\rightarrow n^*}^{\text{c2v}}=0$
    \State Let $\mathcal{S} \leftarrow \mathcal{S}\backslash m^*$
    \For{every $i\in\mathcal{M}(n^*)\backslash m^*$}
        \State Generate and propagate $L_{n^*\rightarrow i}^{\text{v2c}}$ and let $\mathcal{S} \leftarrow \mathcal{S} \cup \{i\}$
        \For{every $j\in\mathcal{N}(i)\backslash n^*$ and $\mathcal{F}(j)=\mathbb{F}$}
            \State Compute $\hat{L}_{i\rightarrow j}^{\text{c2v}}$ and update $R_{i\rightarrow j}^{\text{c2v}}$
        \EndFor
        \State Let $Q(i) \leftarrow \arg\max_{j}\left\{R_{i\rightarrow j}^{\text{c2v}}|j\in\mathcal{N}(i), \mathcal{F}(j)=\mathbb{F}\right\}$
    \EndFor
    \State $\eta \leftarrow \eta + 1$.
    \If{$\eta=E - E_1$}
            \State Set $\mathcal{F}(n)=\mathbb{F}$ for all $n\in \mathcal{O}_N$
            \State Let $\mathcal{S'} \leftarrow \left\{m|d(Q(m))=1,m\in \mathcal{S}\right\}$ and $\mathcal{S} \leftarrow \mathcal{S}\backslash\mathcal{S}'$
            \State Generate and propagate $L_{m\rightarrow n}^\text{c2v}$, then set $R_{m\rightarrow n}^{\text{c2v}}= 0$, where $m \in \mathcal{M}(n)$, $n\in \mathcal{O}_N$ and $d(n)=1$.
    \EndIf

    \If{Stopping condition is not satisfied}
        \State Go to line \ref{ALG:UBPRBPd1_findmaxRSD}
    \EndIf
    \end{algorithmic}
\end{algorithm}

Note that after updating the degree-1 VNs at the end of each iteration,
the unmuted CNs whose $Q(m)$'s are degree-1 VNs should be set muted,
and other CNs would remain their mute/unmute states unchanged.
Such setting is aimed to have CN mute/unmute states
can satisfy the key concept of the UBPRBP decoding--once a CN propagates a message,
it will be set muted and wait for the new incoming V2C information to activate itself. 
\section{Simulation Results}\label{sec:simulation}

In this section,
we compare the proposed algorithms with the original RBP algorithm \cite{RBP}
and the SVNF-RBP algorithm \cite{SVNF},
which is one of the best-performed IDS algorithms in the literature.
Two codes are adopted in our simulations:
one is the (1944,972) code used in IEEE 802.11 (WiFi) standard,
and the other is the (1848,616) codes used for 5G NR,
which is constructed base on the Base Graph (BG) I \cite{NR SPEC} with lift size $Z = 28$.
Note that following the instructions in \cite{NR SPEC},
the NR code is a punctured code generated from a length-1904 mother code,
and the punctured bits are the VNs corresponding to the first two columns in the BG I.
Also, there are 1176 degree-1 VNs in this code.
For all algorithms we present,
one iteration will be counted
after $E$ C2V messages are propagated.
\begin{figure}
  \begin{center}
    \epsfxsize=3.6in \epsffile{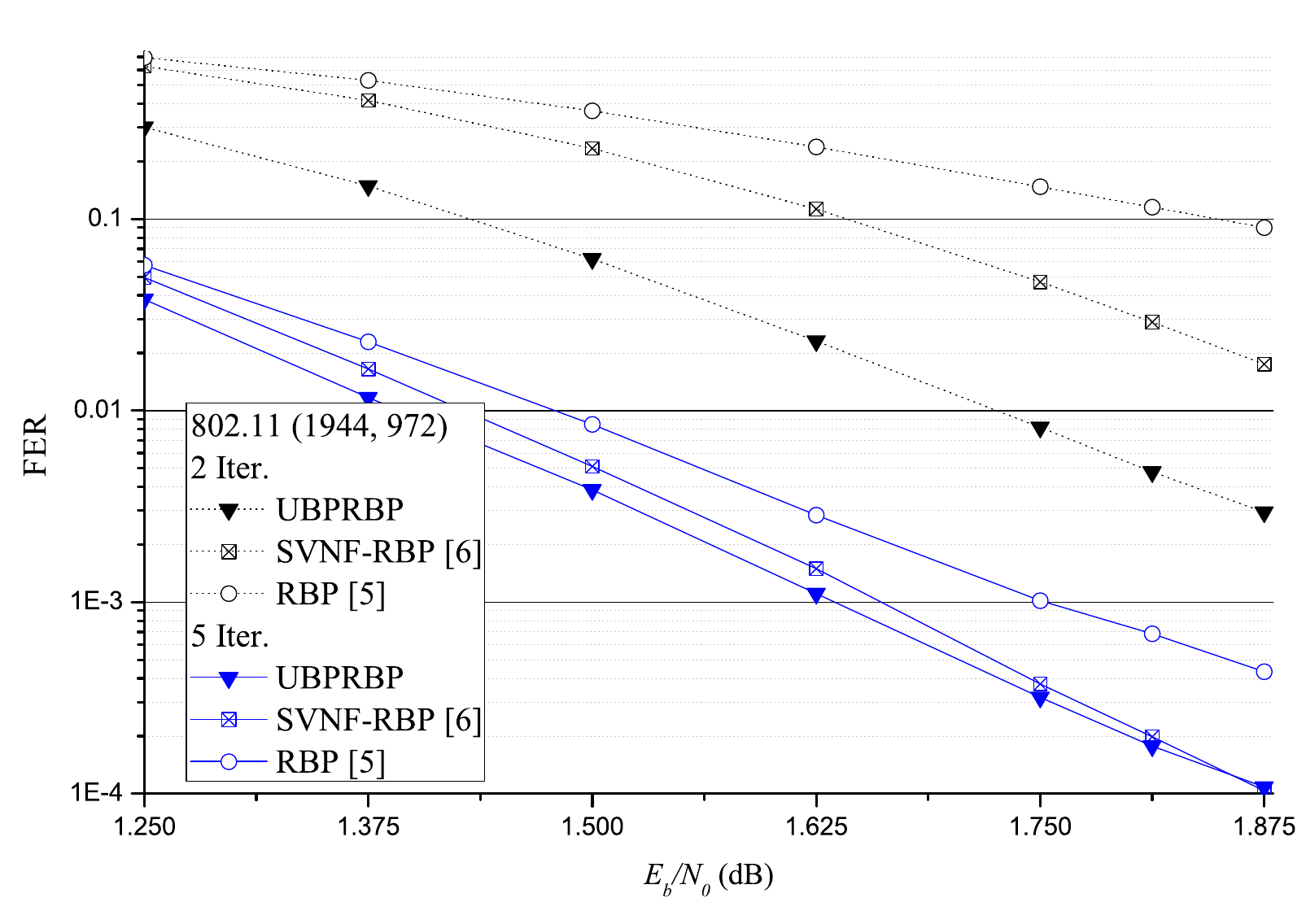}
    \caption{\label{fig:W_SNR_FER}
    FER performance of the RBP [5], SVNF-RBP [6] and UBPRBP algorithms for decoding the (1944,972) WiFi code.}
  \end{center}
\end{figure}

\begin{figure}
  \begin{center}
    \epsfxsize=3.6in \epsffile{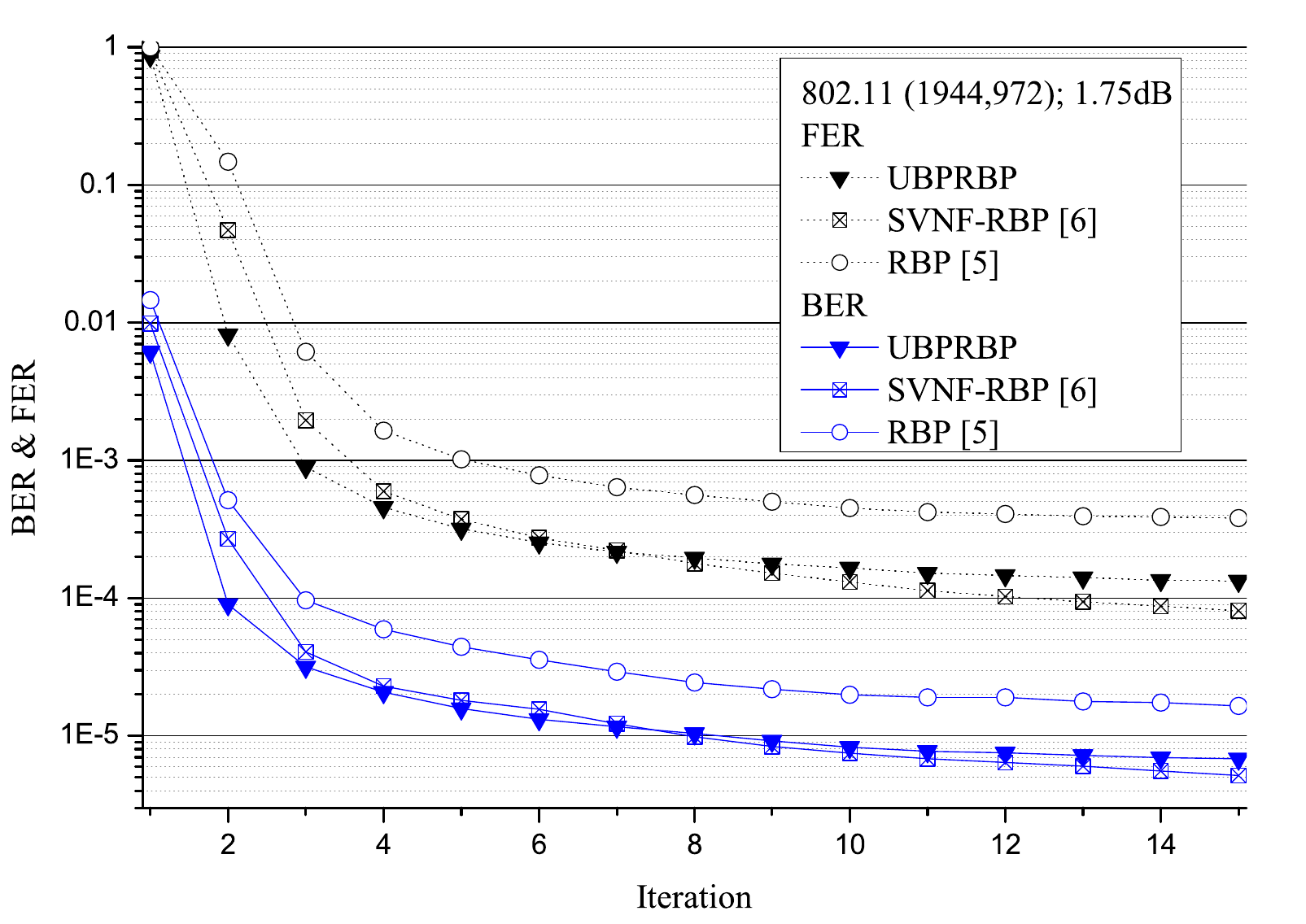}
    \caption{\label{fig:W_ITE_FER}
    BER and FER convergence performance of the RBP [5], SVNF-RBP [6] and UBPRBP algorithms for decoding the (1944,972) WiFi code; SNR=$1.75$ dB}
  \end{center}
\end{figure}

Fig. \ref{fig:W_SNR_FER} shows the frame error rate (FER) performance
of the RBP, SVNF-RBP, and the UBPRBP algorithms in decoding the WiFi code.
The results of 2 and 5 iterations are both presented in the figure.
It can be seen that
the UBPRBP decoder achieves about 0.4 dB and 0.2 dB gains against
the RBP and SVNF-RBP decoder at FER $\approx 10^{-1}$ when the iteration count is 2.
At the 5th iteration,
our UBPRBP algorithm outperforms the RBP decoder about 0.1 dB at
FER $\approx 10^{-3}$.
What presented in Fig. \ref{fig:W_ITE_FER} are the BER and FER convergence performance
of the IDS algorithms in decoding the WiFi code at SNR ($E_b/N_0$)=1.75 dB.
We find that the UBPRBP algorithm is able to outperform the SVNF-RBP decoder
when the iteration count $\leq$ 7;
the SVNF-RBP algorithm can yield better performance
after performing more iterations.

\begin{figure}
  \begin{center}
    \epsfxsize=3.6in \epsffile{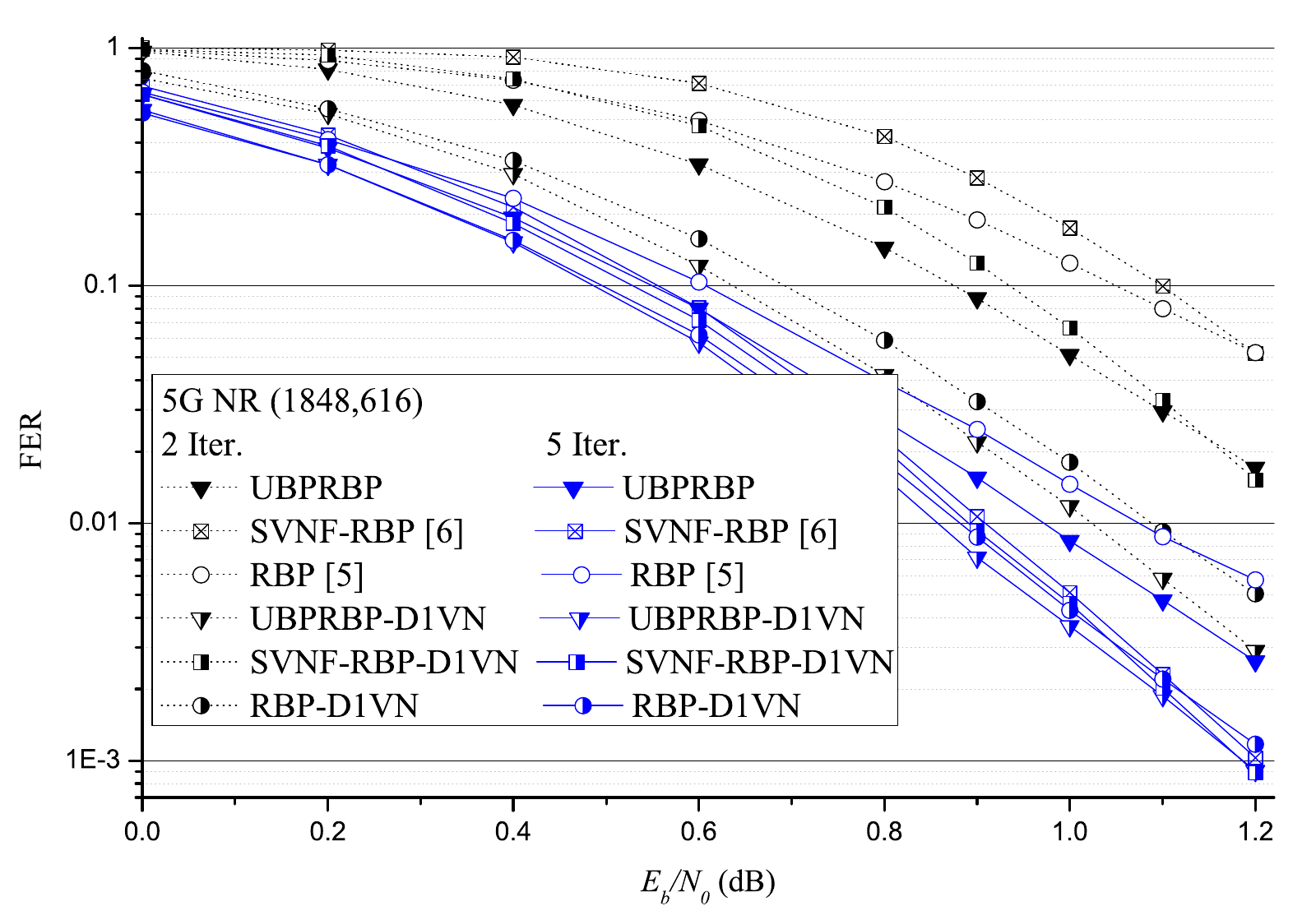}
    \caption{\label{fig:NRd1_SNR_FER}
    FER performance of various RBP-based algorithms for decoding the (1848,616) 5G NR code}
  \end{center}
\end{figure}

\begin{figure}
  \begin{center}
    \epsfxsize=3.6in \epsffile{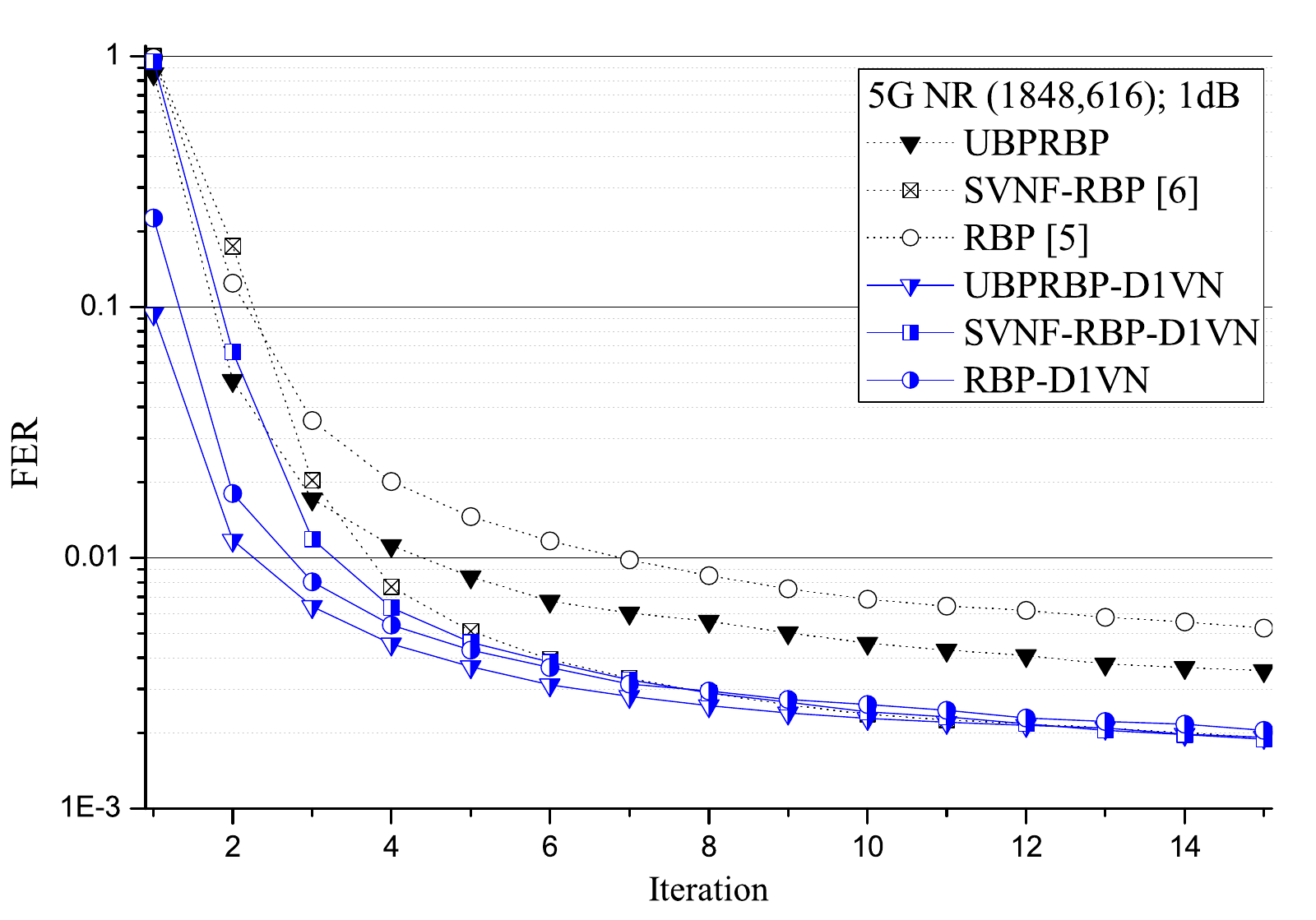}
    \caption{\label{fig:NRd1_ITE_FER}
    FER convergence performance of various RBP-based algorithms for decoding the (1848,616) 5G NR code; SNR=$1$ dB}
  \end{center}
\end{figure}

Fig. \ref{fig:NRd1_SNR_FER} plots the FER performance
of the RBP, SVNF-RBP, and the UBPRBP algorithms in decoding the 5G NR code,
and in  Fig. \ref{fig:NRd1_ITE_FER} we depict the FER convergence performance of these algorithms at SNR=1 dB.
The results similar to that in decoding the WiFi code could be found in the figures:
UBPRBP algorithm can yield superior performance than the SVNF-RBP algorithm in early decoding iterations
but would be interior the SVNF-RBP algorithms as the iteration number increases.
The FER performance curves of combining the RBP, SVNF-RBP, and the UBPRBP algorithms
with our proposed degree-1 VN processing scheme,
which are respectively denoted by RBP-D1VN, SVNF-RBP-D1VN, and UBPRBP-D1VN,
are also presented in Figs. \ref{fig:NRd1_SNR_FER} and \ref{fig:NRd1_ITE_FER}.
We see that with the D1VN processing scheme,
the RBP-D1VN would have a 0.4 dB gain against the original RBP decoder at FER $\approx 10^{-1}$ when the iteration count is 2,
and the SVNF-RBP-D1VN also gains 0.15 dB improvement in comparison with SVNF-RBP decoder.
Also, the UBPRBP-D1VN achieves 0.25 dB gain against the UBPRBP decoder in the same iteration and targeted FER.
Furthermore,
it can be found from Fig. \ref{fig:NRd1_ITE_FER} that
the UBPRBP-D1VN algorithm is able to outperform all the competitors
at all iterations observed
and
the performance of the RBP decoder is impressively improved
after the degree-1 VN processing scheme is adopted--
the RBP-D1VN algorithm could
not only significantly outperform the original RBP algorithm
but surpass the SVNF and even the SVNF-D1VN algorithms.

\begin{figure}
  \begin{center}
    \epsfxsize=3.6in \epsffile{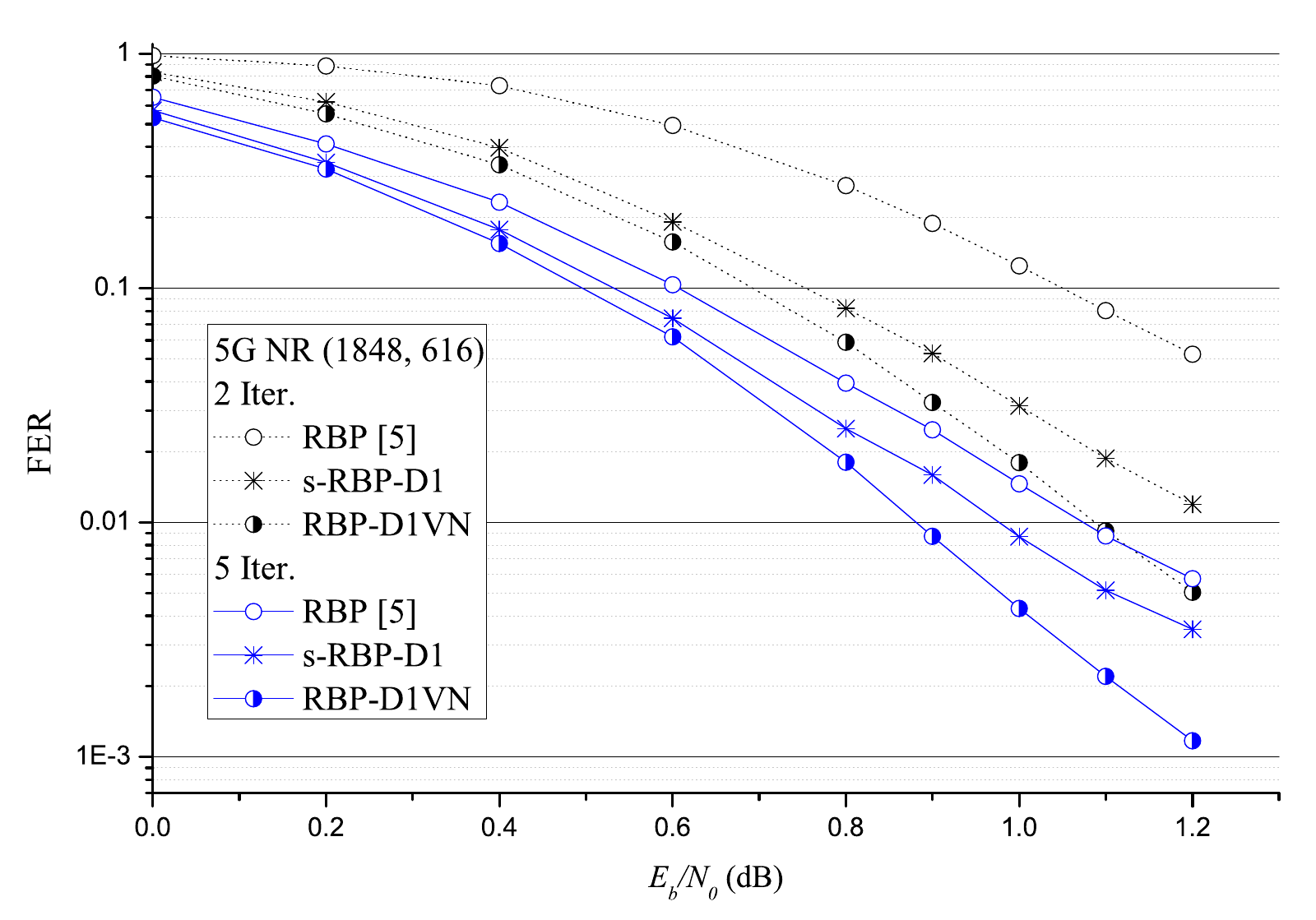}
    \caption{\label{fig:NRd1_RBP_SNR_FER}
    FER performance of RBP [5], s-RBP-D1, and RBP-D1VN algorithms for decoding the (1848,616) 5G NR code}
  \end{center}
\end{figure}

In addition, to verify the effectiveness
of the step used in the D1VN processing scheme that
propagating a substitute C2V message instead of the original selected one as it is associated with a degree-1 VN,
we compare the performance of the original RBP and proposed RBP-D1VN algorithms with
that of a modified RBP which processes the degree-1 VNs with a simple manner.
In this modified RBP, denoted by s-RBP-D1 in the figure,
all the edges connected to degree-1 VNs would not participate the residual comparison during whole decoding process
and will be updated at the end of each iteration.
The simulation results are presented in Fig. \ref{fig:NRd1_RBP_SNR_FER},
and it could be seen although the modified RBP decoder could yield better performance
compared with the original RBP algorithm because it limits the updates of the degree-1 VNs,
the RBP-D1VN algorithm is still the most effective decoder among the three algorithms.

\section{Conclusion}\label{sec:con}

In this paper, we present the UBPRBP algorithm,
which adopts a new IDS strategy, for decoding LDPC codes.
For the UBPRBP algorithm,
we introduce the concept of setting the CNs just delivering C2V messages muted
to mitigate the propagations of unreliable messages.
Experimental results show that
in comparison with the RBP algorithm and the SVNF algorithm,
the proposed UBPRBP algorithm yields superior performance at the early stage of decoding.
The D1VN processing scheme is also presented in this paper.
As the D1VN processing scheme incorporated,
the RBP-based algorithms are able to release the decoding resource
unreasonably occupied by the degree-1 VNs and
hence distribute the resource more effectively
while decoding the 5G NR LDPC codes.
Simulation results indicate that
the RBP-D1VN, SVNF-RBP-D1VN, and UBPRBP-D1VN algorithms would significantly outperform
their corresponding original versions in decoding the 5G NR LDPC codes,
and the UBPRBP-D1VN algorithm is the best-performing one among these three decoders. 


\end{document}